\documentclass{svproc}
\usepackage{url}

\usepackage{amssymb}

\begin{document}
\mainmatter              
\title{Self-interactions of ultralight spinless dark matter to the rescue?}
\titlerunning{Self Interactions}  
%
\author{Gaurav Goswami \inst{1}}
\authorrunning{Gaurav Goswami} 
%
\tocauthor{Ivar Ekeland, Roger Temam, Jeffrey Dean, David Grove,
Craig Chambers, Kim B. Bruce, and Elisa Bertino}
\institute{Ahmedabad University, Ahmedabad, Gujarat 380009, India,\\
\email{gaurav.goswami@ahduni.edu.in}}

\maketitle              

\begin{abstract}
Numerous observations on astrophysical and cosmological scales can be interpreted to mean that, in addition to the familiar kind of matter well described by the standard model of elementary particle physics, there exists Dark Matter (DM). The fundamental properties of the elementary particles which make up the DM e.g. particle mass, spin, couplings etc are currently being observationally constrained. In particular, if DM particles have spin zero, there exist recent constraints which suggest a lower limit on its mass which is often a couple of orders of magnitude larger than $10^{-22}$ eV. In this talk, we will (a) argue that these limits are based on the assumption that the self coupling of the spinless DM particles is negligible, and, (b) show how some of these lower limits will get modified in the presence of incredibly feeble self interactions.

\end{abstract}

\section{Introduction}

Given the evidence in favour of the possibility that Dark Matter (DM, see \cite{Balazs:2024uyj} for review) consists of some particles in one of the extensions of the Standard Model of elementary particle physics, it makes sense to examine the simplest possibility that DM consists of just a single species of (elementary or composite) particles and try to uncover its identity. In other words, one might ask, what is the mass of DM particle, what is its intrinsic spin and what are its couplings to other known particles etc. 
If DM particles are fermions, Tremaine-Gunn bound offers a lower limit on its mass \cite{Balazs:2024uyj} and thus if DM particles are sufficiently light, they can only be Bosons. In this work, we will explore a scenario in which DM is made of a single spinless species and mass as small as possible. Since we know the average density of DM in the Universe, if the individual particle mass is too small, particle number density can be too large. For Bosons, this implies too large occupation numbers and one can achieve coherent classical field states - this is what is called wave Dark Matter \cite{Hui:2021tkt}. Note that this is kind of DM is non-thermal, it is produced in the early Universe by a production mechanism called misalignment mechanism - this leads to background cosmological constraints on  lower limit on particle mass $m \gtrsim 10^{-28}$ eV. Studying the dynamics of first order perturbations teaches us that in such scenarios, since the classical field can't be squashed into too small a region, the matter power spectrum experiences lowering of power on small scales and one can conclude that $m \gtrsim 10^{-23}$ eV. 
 
 We will be concerned with the implications of wave dark matter on nonlinear structures. Some observational constraints seem to rule out $m$ in the range of $10^{-22}$ eV to $10^{-20}$ eV. However, this is based on the assumption that the interactions of DM can be completely ignored and one might then wonder whether couplings of DM could help it in evading these bounds. In particular, for a scalar, since there are also self-couplings, one might ask whether they could help.
 
 \section{Modelling cores of DM halos}

Ultralight dark matter (ULDM) forms cored density profiles (called “solitons”) at the centre of galaxies.  
A configuration of ultra light scalar DM with sufficiently high density is described by classical fields. Since we integrate out rapid oscillations, the resulting equation of motion of the field is classical Schrodinger equation where $|\psi (t, {\bf x})|^2$ is no longer going to be interpreted as the probability to find a single Boson but particle number density (or, mass density, depending on convention \cite{Hui:2021tkt}). In cores of Dark Matter halos, where density is high, this classical field description is valid. For such systems, self gravity is balanced by quantum pressure, the tendency of the condensate wave function to spread. It is useful to compare this with other self-gravitating objects: in main sequence stars and stellar remnants such as White Dwarfs / Neutron Stars, gravity is balanced by thermal pressure and degeneracy pressure respectively.
Mathematically, one can model cores of DM halos by looking for spherically symmetric, nodeless, spatially localised, stationary solutions of Schrodinger-Poisson system of equations which are regular everywhere in space \cite{Chakrabarti:2022owq}. 

\subsection{Self-interactions of ultra-light scalars: theoretical considerations}

How much is the quartic self coupling of scalar ULDM? Is it small enough to be ignorable? This question needs to be answered observationally. 
In order to include the effects of self-interactions of the scalar, one has to solve Gross-Pitaevskii-Poisson (GPP) system of equations. By comparing various terms in the equations, it is possible to estimate the order of magnitude of values of quartic self-coupling which can have observable effects. This order of magnitude of $\lambda$ turns out to be $ \lambda \sim \frac{4m}{M} \frac{L}{m^{-1}},$ where, $m$ is the Boson mass, $M$ is the mass of the core of DM halo (being modelled by the solution of the GPP equation) and $L$ is the spatial extent of the solution i.e. size of the soliton \cite{Chakrabarti:2022owq}. E.g. for $m \sim 10^{-22}$ eV, if the objects of interest (cores of DM halos) have masses ${\cal O}(10^7) M_\odot$ and sizes which are around $10^4$ times the Compton wavelength corresponding to $m \sim 10^{-22}$ eV, the probed $\lambda$ will be ${\cal O}(10^{-92})$.

Before proceeding, it may be useful to compare this probed value with some benchmark values of self coupling expected from theory. For an ultra-light axion with scalar potential $U(\varphi) = \Lambda^4 [1 - \cos (\varphi / f)]$, $\lambda = (\Lambda/f)^4 = (m / f)^2$. Satisfying the relic abundance by misalignment mechanism requires, for $m \sim 10^{-22}$ eV, $f \sim 10^{-96}$ \cite{Chakrabarti:2022owq} which is a few order of magnitude lower than the value which can be probed. One can continue to satisfy the relic abundance constraint and get a larger value of self coupling using clockwork mechanism (see appendix E of \cite{Fox:2023xgx}). 
Before proceeding, we note that even though the axion mass is this small, the axion potential is expected to be stable under radiative corrections since its potential is generated by exponentially suppressed non-perturbative effects. Furthermore, pseudo scalars such as axions easily evade the bounds due to the possibility of mediation of fifth force. 

\subsection{Self-interactions of ultra-light scalars: some observational considerations}

A detailed numerical simulation of wave DM showed that there exists a power law relationship between the DM halo core mass and mass of DM halo itself \cite{Schive:2014dra,Schive:2014hza}. A DM model is no doubt expected to describe observed rotation curves such as those in the Spitzer Photometry \& Accurate Rotation Curves (SPARC) catalogue. Recently there have been arguments ~\cite{Bar:2018acw,Bar:2021kti} that suggest that for ULDM in the mass range $m \in \left[10^{-24}~{\rm eV}, 10^{-20}~{\rm eV}\right]$, both of these things can't be done simultaneously. Since this is a radical claim, one must carefully examine the assumptions behind this claim.
Thus, one can revisit these constraints for ULDM particles with non-negligible quartic self-interactions \cite{Dave:2023wjq} and use a recently obtained soliton-halo relation which takes into account the effect of self-interactions (see \cite{Dave:2023wjq} for details). One then finds that for $m = 10^{-22}~ {\rm eV}$,  the requirement of satisfying both galactic rotation curves as well as SH relations can be fulfilled with repulsive self-coupling $\lambda \sim \mathcal{O}(10^{-90})$.

Another consequence of the wave-like nature of ULDM is the loss of matter from a satellite dwarf galaxy moving in a circular orbit around the centre of a larger host DM halo due to tidal effects. Given the wave-like nature of ULDM, there is a high probability that it will tunnel out of the satellite and if this happens too quickly, then the satellite will not survive over cosmological timescales \cite{Hertzberg:2022vhk}. One can again study the effects of quartic self-interactions \cite{Dave:2023egr} on this problem - and use this to impose constraints on the self-coupling of ULDM. Results of \cite{Dave:2023egr} suggest that for ULDM mass $m = 10^{-22}$ eV, the existence of the Fornax dwarf galaxy necessitates attractive self-interactions with $\lambda \lesssim - 2.12 \times 10^{-91}$.

\section{Discussion}

We thus find that (a) many claims about the ruling out of ULDM in the mass range $10^{-22}$ eV to $10^{-20}$ eV are based on the assumption of negligibly small scalar ULDM self couplings, (b) Observations at the scales of cores of DM halos can probe incredibly small self couplings, (c) taking into account the self interactions of ULDM can save ULDM in the specified mass range from getting ruled out. 
Even the stringent constraints imposed by Lyman-$\alpha$ data can potentially be relaxed if the effects of self-interactions are taken into account (see discussion below eq.~(3.45) in section~3.2.3 in \cite{2168507}, section~5.1.2 in \cite{Ferreira:2020fam}). 
There exist other constraints \cite{Dalal:2022rmp} which may be harder to evade using self coupling alone. Still, this suggests that self interactions of ULDM may be worth taking into account.

{\it Acknowledgement:} I thank Bihag Dave, Koushik Dutta and Sayan Chakrabarti for collaborations on the original works \cite{Chakrabarti:2022owq}, \cite{Dave:2023wjq}, \cite{Dave:2023egr}.

%
%

\end{document}